# A Hybrid Strategy for the Discovery and Design of Photonic Nanostructures

*Zhaocheng Liu[1], Lakshmi Raju[1], Dayu Zhu[1], Wenshan Cai[1,2]\**

[1] School of Electrical and Computer Engineering, Georgia Institute of Technology, Atlanta, Georgia 30332

[2] School of Materials Science and Engineering, Georgia Institute of Technology, Atlanta, Georgia 30332

*Correspondence to: wcai@gatech.edu

**ABSTRACT**: Designing complex physical systems, including photonic structures, is typically a tedious trial-and-error process that requires extensive simulations with iterative sweeps in multi-dimensional parameter space. To circumvent this conventional approach and substantially expedite the discovery and development of photonic nanostructures, here we develop a framework leveraging both a deep generative model and a modified evolution strategy to automate the inverse design of engineered nanophotonic materials. The capacity of the proposed methodology is tested through the application to a case study, where metasurfaces in either continuous or discrete topologies are generated in response to customer-defined spectra at the input. Through a variational autoencoder, all potential patterns of unit nanostructures are encoded into a continuous latent space. An evolution strategy is applied to vectors in the latent space to identify an optimized vector whose nanostructure pattern fulfills the design objective. The evaluation shows that over 95% accuracy can be achieved for all the unit patterns of the nanostructure tested. Our scheme requires no prior knowledge of the geometry of the nanostructure, and, in principle, allows joint optimization of the dimensional parameters. As such, our work represents an efficient, on-demand, and automated approach for the inverse design of photonic structures with subwavelength features.

**INTRODUCTION**

The design and implementation of man-made complex systems have been serving as a driving force in modern science. Such complex physical systems, typically consisting of artificially structured building blocks spanning from the atomic or molecular level to the mesoscale, can be engineered to possess highly unconventional properties or enable distinguishing phenomena not found in their constituent components. As the system complexity grows, it becomes progressively more challenging to propose viable designs and predict resultant behaviors, particularly because theoretical analyses are likely to yield unreliable results due to an increased amount of assumptions and cumulative errors of approximations. Therefore, the trial-and-error method with tedious work and extensive iterations has been prevalently adopted in the design of complex materials and hierarchical structures. With the rapid advances of artificial intelligence (AI) and its related fields, techniques originated from AI are playing a growing role in the interpretation of complex systems with unprecedented accuracy and efficiency [1-4]. In the meantime, novel approaches for the inverse design of complicated materials and chemical compounds using machine learning are gaining growing attentions [5-8]. These AI-based methods have the potential to significantly relieve the computational workloads and expedite the experimental discoveries in research fields pertinent to complex systems.

In the realm of optics, metamaterials and metasurfaces represent complex artificial materials that consist of ordered arrangements of delicately crafted building blocks of deeply subwavelength features [9-11]. These engineered optical media offer unprecedented flexibility in the manipulation of the magnitude, phase delay, and polarization of the interacting light waves [12, 13], and thereby enabling diverse applications in optical imaging, beam steering, light modulation, dispersion engineering, holography, and many more [14-19]. Designing metasurfaces to obtain aimed optical properties, however, is usually a tedious trial-and-error process that requires a wealth of knowledge and experience in the optical domain. The process often starts with an empirical guess from an experienced researcher, followed by iterative refinement of all geometrical and material parameters through extensive numerical modulations. In many cases, the enormous degrees of freedom of a meta-structure can impede the optimization process in the multi-dimensional parameter space, even after considerable rounds of iterative parametric sweeps.

As for the inverse design of optical structures, two approaches are notably investigated in the field of AI – the evolutionary algorithm (EA) [20-22], which is in the domain of optimization [23, 24], and traditional neural networks (NN) [25-27], which are classified as a data-driven method. Both schemes are capable of optimizing the parameters of a certain structure, either by the evolution of the parameter list or through the backpropagation from a trained neural network. While these two methods have led to a number of successful demonstrations, they both suffer from stringent constraints on the diversity of structures, which are typically limited to a fixed shape with a couple of adjustable parameters. This challenge can be alleviated by leveraging a generative adversarial network (GAN), which identifies an optimum structure from a geometrical dataset based on customized design criteria (e.g., optical spectra) at the input [28, 29]. However, this GAN-based approach still requires some prior knowledge of the general classes of the geometry in order to avoid slow convergence. In addition, given the fact that a vast quantity of data is not readily available from the simulations, a network, even trained with gigantic data entries, can only solve specific design problems, strictly correlated to the data used in the training process of the network.

To automate the inverse design of optical structures with minimal intervention of human, and to discover arbitrarily-shaped photonic building blocks with minimal predefined restrictions, here we introduce a fast, robust, and generalizable strategy, consolidating a deep generative model and a modified evolution strategy (ES). The designing process only takes customer-defined, on-demand physical properties as the input, without any other prior knowledge. Moreover, the framework is capable of generating photonic patterns with nanoscale precision, in either continuous or discrete topology. A generative model, specifically a variational autoencoder (VAE) [30] in this case, is exploited to encode all potential structures into a latent space, and the modified ES algorithm is applied to the latent vectors to identify an optimum pattern whose physical properties fit the target ones at the input with minimized discrepancy. The proposed framework does not require a trained neural network simulator to perform the backpropagation as used in inverse design approaches reported thus far, and a traditional physical simulator can be incorporated when the training data for the neural network is computationally expensive to acquire. However, a neural network simulator can drastically decrease the searching time from days to seconds as will be presented in this paper. Our results show that an average accuracy of above 95% can be achieved for all the test samples, each taking no longer than 5 seconds. Finally, the scalability of the framework enables our approach to be implementable to the inverse design of complicated photonic structures with multiple dimensional parameters concurrently optimized.

**RESULTS AND DISCUSSION**

**The VAE-ES framework**. In order to utilize the evolution strategy for the discovery of desired photonic structures with an arbitrary pattern of the unit cell, we first construct a compact representation of all possible candidates of the geometry. Derived from deep neural networks, this compact representation can be computed by encoding the geometric data into a latent space with the help of a deep generative model, such as a generative adversarial network (GAN) or a variational autoencoder (VAE). The latter is adopted here for the representation of geometrical shapes of photonic nanostructures. **Figure 1**(a) represents a vanilla VAE used to compute the latent space of all unit-cell patterns in this work. The encoder transforms the input geometric data into mean vectors $\mu$ and standard deviation vectors $\sigma$, and latent vectors $v$ are sampled from the Gaussian distribution parametrized by $\mu$ and $\sigma$. The decoder then reconstructs $v$ back to the geometric information. When the training of this VAE is completed, the decoder can be operated as a geometric data generator as indicated in Fig. 1(b), so that when fed with a randomly sampled vector $v$, the corresponding pattern of the nanostructure $s$ can be reconstructed.

With the decoder of a trained VAE as a generator ($G$), we can optimize the pattern of the nanostructure in the unit cell in response to specific design objectives, denoted by $q$, by applying the ES to the latent space of the generator. As illustrated in **Figure 2**, the framework is built upon, but differs drastically from, the traditional ES, particularly in the evaluation of the fitness score. In the implementation process, a population of all individuals is first initialized. Each individual contains two random vectors – a latent vector $v$ and a mutation strength $m$. After that, the fitness scores $r$ of the population are evaluated based on a simulation process ($S$) and a fitness function ($F$). To achieve this, the latent vector $v$ is decoded to an image of photonic structure using the generator, and simulation for each individual $v$ is carried out. The simulated results $\hat{q} = S(G(v))$ are compared against the physical properties of interest at the input, yielding scores $r = F(\hat{q}, q)$ representing the agreement between the properties of the individuals and the input one. After the evaluation, $\mu$ best individuals with the highest scores are selected as parents ($P$), and their reproduction leads to the next generation of children ($Q$) with a total number of individuals $\lambda$. The latent vectors $v$ of children are created via conventional crossover, which randomly exchanges segment of vectors of the two parents, and interpolation, which means interpolating the vectors of the two parents with random weights. The mutation strength $m$ of a child is derived in the same manner as $v$. The $\lambda$ reproduced individuals together with the $\mu$ parents construct the next generation, which is known as the ($\mu+\lambda$) strategy. In the mutation stage, noises sampled from the normal distributions with mean 0 and standard deviation $m$ are add to the latent vector $v$. The mutated population then goes through the iterative process of evaluation, reproduction, and mutation until a satisfactory individual with an optimized score is identified. We hereby name the framework the VAE-ES framework, as it represents a consolidation of two separate AI algorithms, the VAE and the ES.

**Evaluation of the VAE-ES Strategy.** To evaluate the performance of the proposed strategy for the inverse design of photonic structures, the VAE-ES framework is applied to the design of the unit-cell pattern $s$ of a metasurface that corresponds to customer-defined optical spectra $T$ fed at the input. A VAE is trained with a pyramid convolutional network architecture[31] using a geometric dataset that contains more than 10,000 random patterns of various shapes. The classes of all geometric data include, but not limited to, ellipse, rectangle, polygon, cross, random union/intersection of other classes, and false reconstructed samples by-produced from the training process of the VAE. Each geometric data point is represented as a binary pixel image with a size of 64 by 64, in which 1 stands for the constituent material and 0 for the air, and the encoded vector $v$ of the images by the VAE has the dimension of 10. To accelerate the searching speed, we implement a neural network simulator to approximate the spectral behavior of the metasurface with a specific unit-cell pattern. The simulator accepts the image of a nanostructure and approximates the four components of the transmittance $T_{ij}$ of the metasurface, where $i$ and $j$ indicate the polarization directions for the incidence and the detection, respectively. The frequency range of spectra under investigation is set to be from 1.7 THz to 6 THz, which corresponds to 500 nm – 1.8 μm in the wavelength domain. The default lattice period and the thickness of the patterned layer are set to be 380 nm and 50 nm, respectively. Gold is chosen to form the single-layered nanostructure, which situates on a glass substrate. It is worth noting that the neural network simulator used here is developed to expedite the computing process, and can be replaced by any physical simulation method such as the finite element method (FEM) and the finite-difference time-domain (FDTD). We also note that the restrictions with regard to the frequency range, the lattice period, the thickness of the patterned layer, and the type of constituent materials are predefined by the simulator. Trained simulators of other metrics would allow the framework to inversely design photonic structures under other settings.

In the next step, we define the fitness function and illustrate the experiment settings. We denote the patterns identified from the framework as $s'$, the spectra approximated by the neural network simulator as $\hat{T}$, and the transmittance of $s'$ computed by the FEM simulation as $T'$. As the objective is to minimize the $L_2$ and/or $L_\infty$ norm of the difference between the input spectra and the approximated transmittance, a reasonable choice of the fitness function can be written as

$$F(T,\hat{T}) = -\|T - \hat{T}\|_2 - \beta \max(T - \hat{T}), \qquad (1)$$

where $\beta$ is the weight balancing the two norms. Both norms are included to stabilize the searching process and to enforce similar numerical features between $\hat{T}$ and $T$. Setting $\beta = N$, where $N$ is the dimension of the vector $T$, allows the values of the two norms to be on the same order of magnitude. Since we do not rely on the backpropagation algorithm in the ES, the fitness function can be nondifferentiable. In the experiment, the population size is set to 50, and the children size of each new generation is from 50 to 100. Only individuals with top 25 scores are selected as parents of the next generation. The maximum iterations of the algorithms are set to 200, which on average takes 5 seconds for each set of the input spectra with the acceleration of the simulator. Based on our extensive testing, most of the searching practice will stop as an optimized score is reached within 50 generations. It should also be noted that the parent size and the children size are not restrictive to the values we described above. Changing the parameters of the ES causes a very minor impact on the final performance of the design. To quantitatively evaluate the performance of our approach, the average accuracy of each identified pattern *s'* is defined as:

$$a(T_{sim}, T) = 1 - \frac{1}{f_2 - f_1} \int_{f_1}^{f_2} |T_{sim} - T| df, \qquad (2)$$

where $f_1$ and $f_2$ are the frequency bounds of the input spectra, and $T_{sim} = \hat{T}$ or *T'* represents the accuracy of *s'* calculated with the simulated results from the neural network simulator or from the FEM simulator.

We demonstrate the capability and accuracy of our framework by applying it to the inverse design of photonic structures based on the input spectra of randomly selected, actual metasurfaces. This ensures that a reasonable solution must exist for any given input spectra of this kind. In our experiment, 500 random test patterns of metasurfaces *s* from all classes of geometry are selected, and FEM simulations are performed to obtain the spectra *T* of these metasurfaces, which are subsequently fed to the framework as the target spectra. Our algorithm is tested by identifying the corresponding pattern *s'* for the input spectra *T* and evaluating the average accuracy of the simulated spectra of *s'*. Statistically, the expected $a(\hat{T}, T)$ and $a(T', T)$ of identified *s'* are above 98% and 95%, respectively, and the difference stems from the error associated with the neural network simulator. **Figure 3** presents twelve examples of the input spectra *T* (solid lines) and FEM simulated spectra *T'* (dashed lines) in conjunction with the corresponding patterns *s* and *s'* in the unit cell. Our approach captures the prominent features of the input transmittance and generates nanostructures with minimal discrepancy between *T* and *T'*. As the ES algorithm only considers the distance between *T* and $\hat{T}$ as the optimization criterion, any shapes, similar to or distinct from $s$, can be identified as an optimal solution only if the score $F(T, \hat{T})$ is maximized. This feature also allows us to unearth different possible topologies in response to a given $T$ in multiple runs.

In the next part, we will reveal the capacity of the proposed framework for the inverse design of photonic structures, based on arbitrary, user-defined input spectra. Two sets of experiments for such on-demand design have been performed, and the results are shown in **Figure 4**. In Figures 4(a) to 4(h), the input spectra $T_{xx}$ and $T_{yy}$ are set to be two randomly chosen, Gaussian-shaped curves, and $T_{xy}$ and $T_{yx}$ are set to zero throughout the frequency range of interest. The results of the FEM-simulated spectra of the identified patterns (dashed lines) faithfully match the input spectra in terms of both the spectral location and the bandwidth. In Figure 4(i) to 4(p), we set the input $T_{xx}$ as a notch filter without any specifications in regard to the $T_{yy}$, $T_{xy}$ and $T_{yx}$ components. We note that within the family of single-layered metallic metasurfaces, it is not possible to identify a metasurface that perfectly replicates ultrasharp spectral features such as the steep cut-off slopes. However, the VAE-ES framework is able to generate a nanostructure that accurately replicates the primary features of the desired filter, including the central frequency and bandwidth of the stop-band. In the general case when no pattern exists in correspondence to a given desired spectra at the input, the framework is guaranteed to generate a nanostructure with minimized discrepancy between its spectra and the input ones. We also note that although the transmittance spectra are set as the design objective in these experiments, any photonic responses such as the diffraction behavior, optical chirality, and field localization can be used as the intended design criteria without further adjustment of our framework.

**Generalization of the VAE-ES framework**. As shown from the examples presented above, the VAE-ES developed here is capable of generating patterns of single-layered metasurfaces with prescribed properties, but certain dimensional and physical parameters are preset. While limited by the availability of data, here we briefly commend on how the proposed methodology can be adapted to more complicated photonic structures that consist of multi-layered or three-dimensional build blocks. For example, by expanding the latent vectors with additional parameters, our VAE-ES approach can be applied to the optimization of an extended parameter space without any adjustment. As presented in **Figure 5**(a), when the latent vector is expanded to include extra parameters such as the thickness, the lattice period, and the materials, the VAE-ES framework will be able to identify the nanostructured pattern while concurrently seeking the optimum parameters. Figure 5(b) illustrates an encoding method for the inverse design of multi-layered metasurfaces. The encoded vectors of the layers are concatenated, with an additional parameter representing the distance between adjacent layers. There are other effective encoding methods for more complicated photonic structures, especially when geometrical features along all dimensions are involved, although a complete discussion of these schemes would be beyond the scope of the current work. The common characteristic of these encoding methods is to exploit a well-trained VAE or GAN to offer a compact representation of all possible candidate structures.

**CONCLUSION**

In summation, we have demonstrated a robust, efficient, and generalizable framework consolidating the evolution strategy and a generative model to automate the inverse design of photonic nanostructures for user-defined design objectives. The capacity of the proposed framework is tested through design practices of metasurfaces in response to on-demand spectral properties. Without prior knowledge of the geometry of the candidate patterns, the VAE-ES framework automatically generates the optimum structure, in continuous or discrete topology, of the metasurface, with a fidelity significantly higher than those of current NN-based methods. Moreover, our approach allows flexible substitution of constituent modules in the algorithm. For example, the user can choose to replace the generator with either a VAE or a GAN trained on the dataset of interest, and determine the favorite simulation method such as a neural network (to expedite the search speed) or the FEM (to improve the reliability of the design). Unlike other NN-based approaches, the fitness function in our framework is not restricted to differentiable functions. As such, the fitness function can be formulated to possess a large degree of freedom to accommodate arbitrary physical objectives, an important feature shared among traditional optimization methods.

We envision that the performance of the VAE-ES framework can be consistently boosted by advanced generative models and evolutionary algorithms. Thanks to the flexible formulation of the fitness function and the scalability of the encoding method for photonic structures, the strategy developed in this work can be adapted for the design and implementation of complex photonic materials and devices for prescribed optical properties and diverse light manipulation capacity. Without any working knowledge of optics, our approach can be readily implemented to speed up the designing process of large-scale photonic devices such as meta-holograms and photonic crystals. Finally, beyond the field of optics and photonics, we envisage a generic utilization of our framework in other disciplines where people can be, to a certain extent, spared from conventional trial-and-error practices when designing complex material systems.

**Figures and Figure Captions**

**Figure 1**

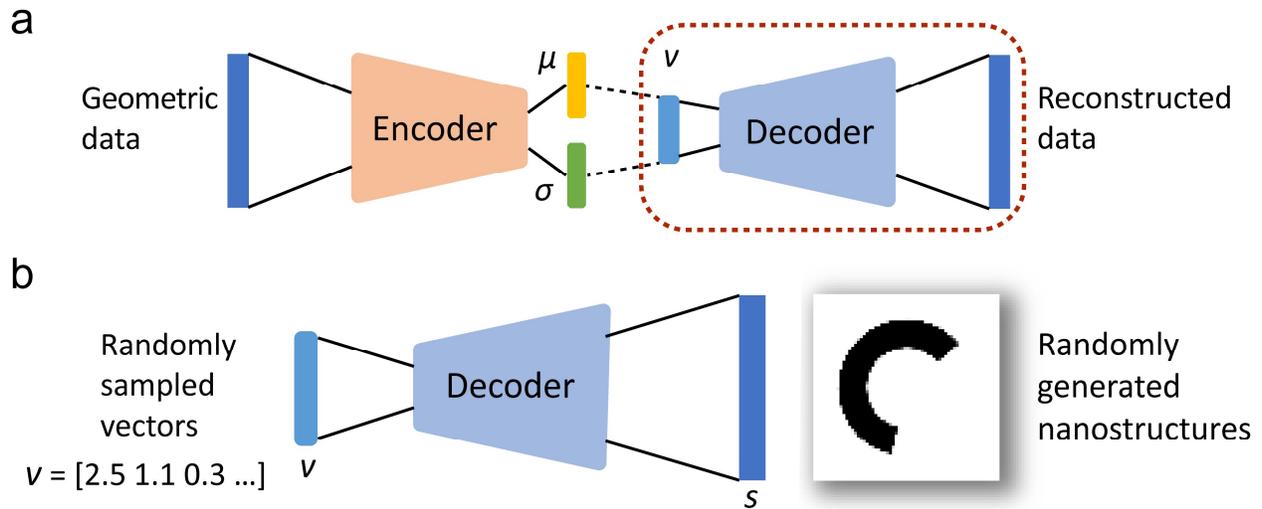

**Figure 1: Basic architecture of a VAE implemented in the framework.** (a) The training process of the VAE. The encoder accepts the geometric data and produces two parametric vectors $\mu$ and $\sigma$. Random vectors *v* are samples from the normal distribution with the mean and the standard deviation defined by $\mu$ and $\sigma$. The decoder then reconstructs vectors *v* to images of nanostructures. The VAE encodes the geometric data into a compact latent space where the ES can be applied efficiently. (b) After the training, the decoder encircled in (a) can be separated and treated as a generator of geometric data. It transforms randomly sampled vectors *v* to their correspondent nanostructures.

Figure 2

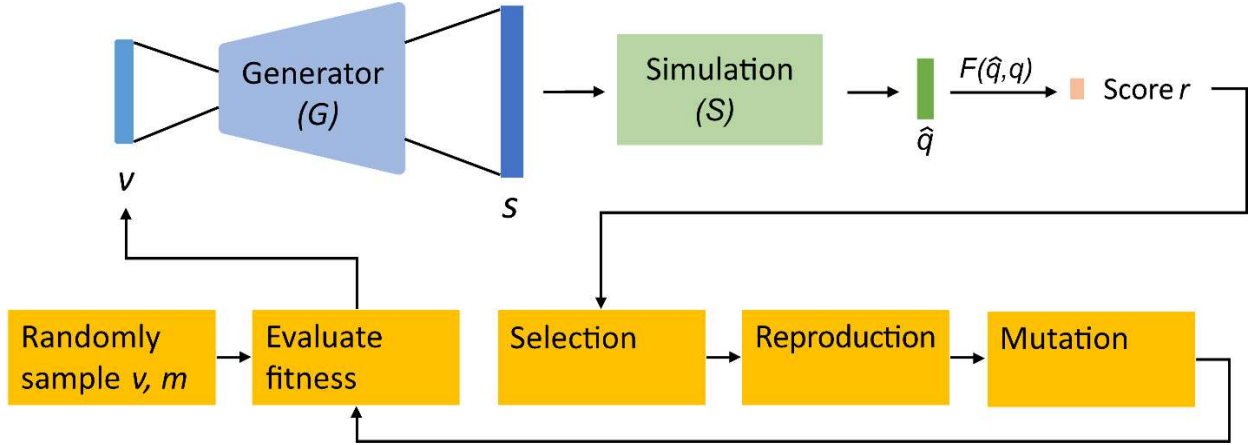

**Figure 2: Flowchart of the VAE-ES framework.** The developed approach employs a traditional ES algorithm with a loop of evaluation of fitness, selection, reproduction, and mutation. When evaluating the finesses of all individual vectors $v$, the generator ($G$) is utilized to reconstruct the nanostructure of each $v$. Simulation process ($S$) is then followed to produce the simulation results $\hat{q}$. The fitness function compares $\hat{q}$ to the desired physical properties $q$, and results in a score $r$ that represents the agreement of $\hat{q}$ and $q$. This score is an indication of the fitness of $v$. Note that the simulation can be carried out by either a neural network approximation or a real physical simulation such as FEM.

**Figure 3**

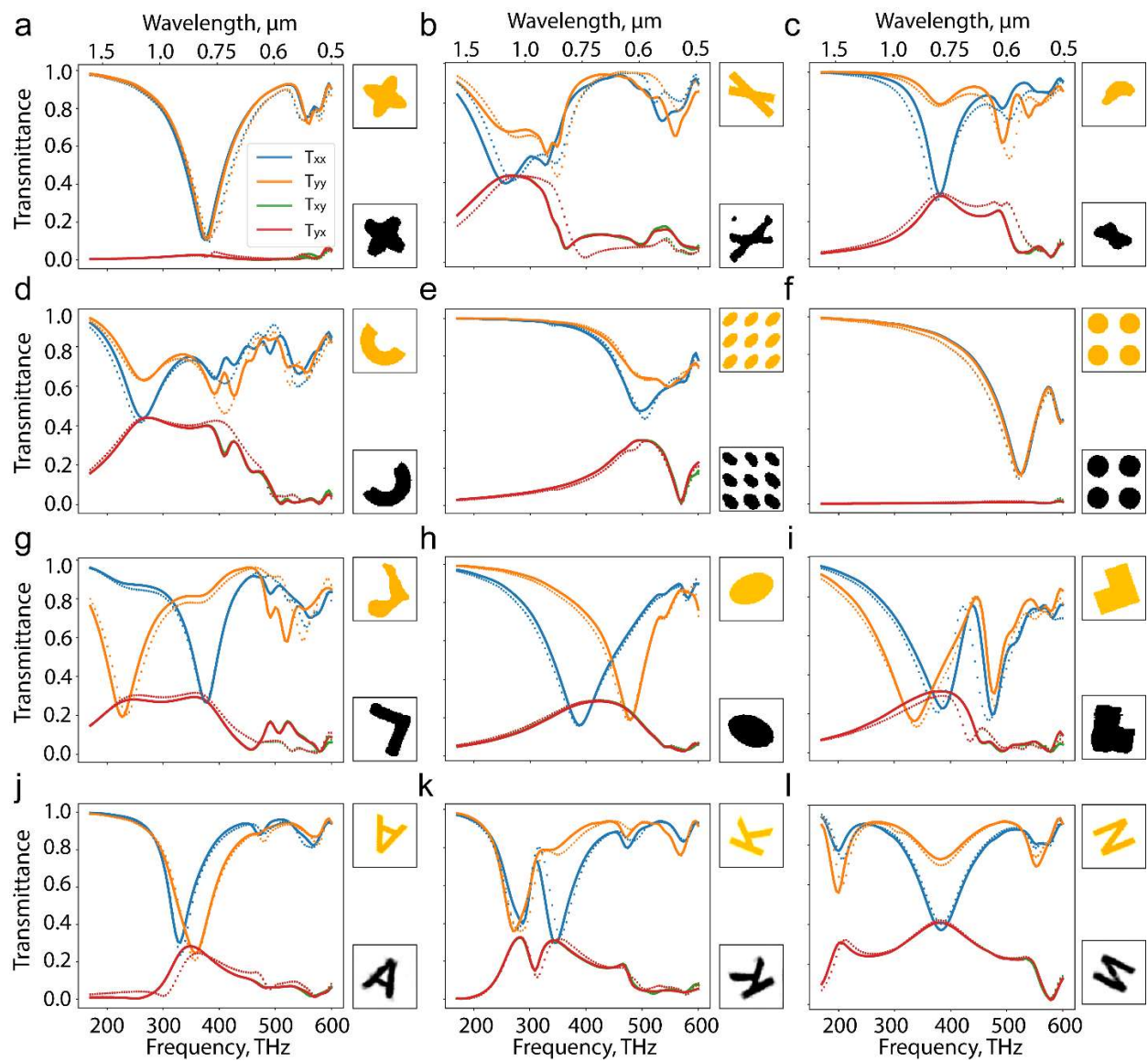

**Figure 3: Samples of results using spectra of existing nanostructures as the input.** (a) – (l) The desired spectra $T_{xx}$, $T_{yy}$, $T_{xy}$ and $T_{yx}$ drawn from a series of test patterns are shown in solid lines, while the FEM simulated spectra $T'$ of each generated pattern are shown in dashed lines. The test patterns *s* and discovered patterns *s'* are presented on the right of each figure in orange and black, respectively.

**Figure 4**

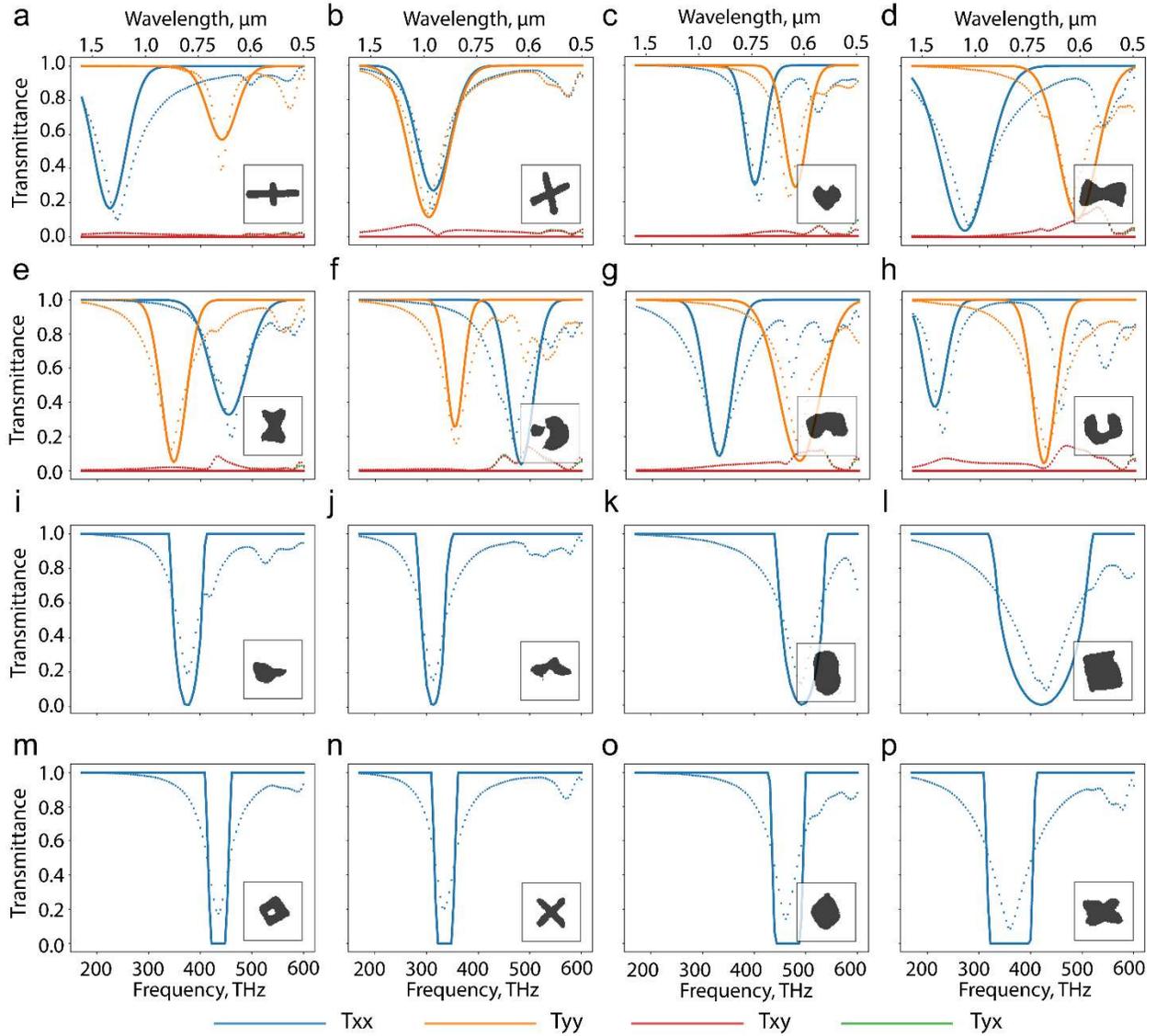

**Figure 4: Samples of on-demand inverse design of metasurfaces.** (a) – (h) The desired spectra $T_{xx}$ and $T_{yy}$ at the input, shown as the solid lines, are two randomly generated Gaussian-like curves, and the $T_{xy}$ and $T_{yx}$ are zeros across the frequency range of interest. The generated patterns in the unit cell are depicted in the lower right corner of each panel, and the FEM-simulated spectra of the resultant nanostructures are represented in dashed lines. (i) – (p) Inverse design of notch filters, where the desired spectrum $T_{xx}$ at the input has a band-stop transmission feature with specific central frequency and bandwidth. All these examples demonstrate the effectiveness of the framework, which is able to generate nanostructures that resemble the on-demand spectra fed at the input, and faithfully replicate major features in terms of the spectral location and the bandwidth.

**Figure 5**

a. Encoding method for single-layered metasurfaces

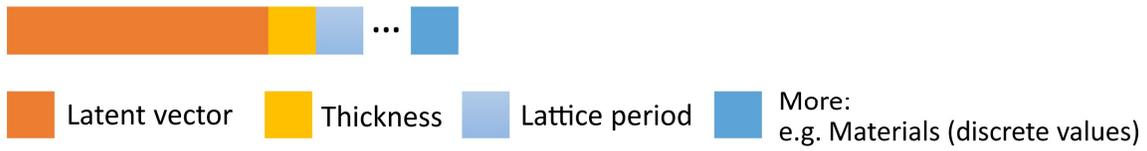

b. Encoding method for multi-layered metasurfaces

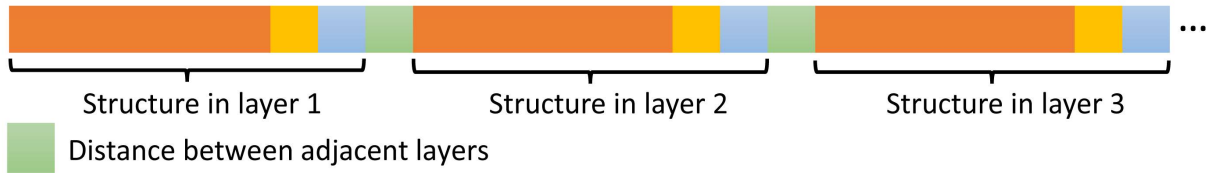

**Figure 5: Encoding methods for the general inverse design of metasurfaces**. (a) Encoding a nanostructure in a unit cell with additional parameters, such as the thickness of the patterned layer, the lattice constant of the unit cell, and the constituent materials encoded in discrete values. (b) An encoding method for multi-layered metasurfaces. A series of encoded vectors, representing nanostructures in different layers, can be concatenated as a new vector for the inverse design of multi-layered metasurfaces. The distance between adjacent layers can also be included as a parameter to be optimized.

18